\documentclass[manuscript]{aastex}

\slugcomment{to appear in ApJ}

\shorttitle{Water ices and hydrocarbons in the central parsec}
\shortauthors{Moultaka et al.}

\begin{document}


\title{"Ice cubes" in the center of the Milky Way - \\
    Water ice and hydrocarbons in the central parsec\altaffilmark{1} } %thanks{Resulting from ESO VLT observations of program ID numbers 71.C-0192A and 077.C-0286A}}


\author{J. Moultaka}%\altaffilmark{1} and Ivan R. King\altaffilmark{1}}
\affil{Universit\'e de Toulouse; UPS-OMP; IRAP; Toulouse, France\\
CNRS; IRAP; 14, avenue Edouard Belin, F-31400 Toulouse, France}
\email{jihane.moultaka@irap.omp.eu}
\author{A. Eckart} % Biemesderfer\altaffilmark{4,5}}
\affil{I Physikalisches Institut, Universit\"at zu K\"oln, Z\"ulpicher Str. 77, D-50937 K\"oln, Germany}
\email{eckart@ph1.uni-koeln.de}

\and

\author{ K. Mu$\check{\bf z}$i\'c}
\affil{European Southern Observatory, Alonso de C\'ordova 3107, Casilla 19, Santiago, 19001, Chile}


\altaffiltext{1}{Resulting from ESO VLT observations of program ID numbers
71.C-0192A and 077.C-0286A}
\begin{abstract}
The close environment of the central supermassive black hole of our Galaxy is studied thoroughly since decades in order to shed light on the behavior of the central regions of galaxies in general and of active galaxies in particular. The Galactic Center has shown a wealth of structures on different scales with a complicated mixture of early- and late-type stars, ionized and molecular gas, dust and winds.\\
Here we aim at studying the distribution of water ices and hydrocarbons in the central parsec as well as along the line of sight. This study is made possible thanks to L-band spectroscopy. This spectral band, from 2.8 to 4.2$\mu m$, hosts important signatures of the circumstellar medium and interstellar dense and diffuse media among which deep absorption features are attributed to water ices and hydrocarbons.\\
We observed the Galactic Center in the L-band of ISAAC spectrograph located on UT1/VLT ESO telescope. By mapping the central half parsec using 27 slit positions, we were able to build the first data cube of the region in this wavelength domain. Thanks to a calibrator spectrum of the foreground extinction in the L-band derived in a previous paper, we corrected our data cube for the line of sight extinction and validated our calibrator spectrum. The data show that a residual absorption due to water ices and hydrocarbons is present in the corrected data cube. This suggests that the features are produced in the local environment of the Galactic center implying very low temperatures well below 80K. This is in agreement with our finding of local CO ices in the central parsec described in Moultaka et al. (2015).
\end{abstract}


\keywords{Galaxy: center - galaxies: nuclei - infrared: ISM}



\section{Introduction}

  The proximity of our Galactic Center (hereafter GC), located at only 8~kpc makes it the best example of a central galactic region that can be studied in detail and at a high spatial resolution. The presence of a supermassive black hole (associated with the radio source SgrA*) of $\sim 4\, 10^6$ M$_\odot$ in the dynamical center of our Galaxy has been established during the past two decades 
(Eckart \& Genzel 1996, Sch\"odel et al. 2002, Ghez et al. 2003). Since then, numerous works aim at studying the activity of this black hole as well as its influence on the direct environment of gas, dust and stars.\\ 
Understanding the composition and the working of this environment is a unique opportunity to shed light on the physics at work in the central regions of galaxies in general, in active galaxies in particular, and at the same time to improve our comprehension of galaxy evolution.\\

The central ten parsecs of the Galaxy show a very rich environment made of dust, neutral and ionized gas where bright infrared sources are observed and identified as late-type, early-type or dust-embedded stars. A molecular component called the Circumnuclear Disk (CND) in the shape of a torus with an outer edge located at about 7~pc from SgrA* is observed in HCN emission map by G\"usten et al (1987). It is composed of dust clouds and molecular dense and warm gas orbiting and falling towards the center from its inner edge at 1.5 pc (Lutz et al. 1996; de Graauw et al. 1996; Gerakines et al. 1999, Lau et al. 2013).
Inside the one parsec diameter cavity surrounded by the CND, streamers of ionized gas (also called the mini-spiral structure) are likely bringing material to the center (Yusef-Zadeh \& Morris 1987, Zhao et al. 2010). \\

In this work, we use the ESO/VLT ISAAC spectrograph in order to study the spatial distribution of water ices and hydrocarbons in the half parsec region around SgrA$^{\star}$.
The L-band spectral range of ISAAC (from 2.8 to 4.2$\mu$m) hosts a broad absorption feature at $\sim 3\mu$m, due to the O-H stretching vibration of water ice mantles condensed on dust grains present in the dense ISM. This feature has been often observed towards Young Stellar Objects (YSOs) (e.g. Ishii et al. 1998, Brooke et al. 1999) and is usually associated with star forming regions. \\
Water vapor is observed in different objects in the universe, from exo-planetary disks and comets to star forming regions and galaxies (Hartogh et al. 2011a,b, Hogerheijde et al. 2011, van Dishoeck et al. 2011, van der Werf et al. 2010, Weiss et al. 2010). It was recently observed by Herschel Space Observatory (Caselli et al. 2012), for the first time, in the pre-stellar core L1544 and is shown to be produced by far-UV irradiation of icy dust mantles induced by cosmic-rays.\\
Two absorption features can also be distinguished in the L-band spectra of the diffuse medium at 3.4$\mu$m and 3.48$\mu$m. They are attributed to C-H stretching vibrations in the CH$_2$ and CH$_3$ groups of aliphatic hydrocarbons. \\

The previously mentioned absorption features have been first observed toward the Galactic Center by ISO/SWS (Lutz et al 1996). They were also observed from the ground by a number of groups (Hoyle \& Wickramasinghe 1980a,b, Jones et al. 1983, Butchart et al. 1986, McFadzean et al. 1989, Sandford et al. 1991, Wada et al. 1991, Pendleton et al. 1994, Gao et al. 2010, Chiar et al. 2002, Mennella et al. 2003). All previous studies concluded that these features arise from the foreground molecular clouds given the fact that no such compounds can survive in the harsh and hot environment of the Galactic Center. Moreover, the lack of correlation between the strength of the water ice and hydrocarbon features seen in the direction of the GC is interpreted as being evidence of distinct interstellar medium carriers for these two compounds (Sandford et al. 1991).\\
 
Moultaka et al. (2004, 2005) observed a number of Galactic Center bright infrared sources with ISAAC spectrograph (ESO/VLT telescope). We found, that, unlike the previous statements, part of the absorption features may arise from the local medium of the central parsec. This result was obtained by calibrating the foreground extinction to the GC in the spectral band and correcting the observed spectra for the extinction. This is also in agreement with our recent findings of CO ices in the local environment of the GC (Moultaka et al. 2009 and 2015). \\
We also found a correlation between the strength of the water ice feature and that of the hydrocarbon feature suggesting a mixture of the dense and diffuse interstellar media.\\
In addition to these absorption features, hydrogen and helium emission lines are also observed in the L-band spectra toward the Galactic Center. These are signatures of the ionized gas and Wolf-Rayet stars that are identified in the region (e.g. Krabbe et al. 1995, Najarro et al. 1997, Buchholz et al. 2009). \\

 Here, we present the first data cube of the central half parsec of the Galaxy obtained in the wavelength range from 2.8 to 4.2$\mu m$ and corrected for the foreground extinction. We show for the first time, the intrinsic spectra of the local environment of the central half parsec in this spectral domain and the spatial distribution of the spectral features. In section 2, we describe the observations and the data reduction that allowed us to build the data cube of the central half parsec. In section 3, we explain the method that we used to calibrate the foreground extinction in the L-band and provide strong arguments validating the method. We analyze the resulting data cube in section 4 and confirm the correctness of the foreground extinction calibrator spectrum with additional arguments. Finally, a summary and conclusion are given in section~5. 

\section{Observations and data reduction}

We have used the ISAAC spectrograph located on the UT1 ESO/VLT telescope (program 077.C-0286A) in order to complete the mapping of the central half parsec of the Galaxy in the L-band. 
With the same spectrograph and instrument setup (Moultaka et al. 2005) we had already started this project in observing period 71.
To this end, we used 19 slit positions placed parallel to each other with an angular offset on the sky of 0.6" (corresponding to the slit width) in addition to the 8 positions previously observed in period 71 (see Fig.~\ref{Lband}). The seeing varied from 0.6" to 1.4". We used the long wavelength (LWS3) and low resolution (LR) mode with the SL filter covering the wavelength range from 2.7$\mu m$ to 4.2$\mu m$. Combined with a 0.6" slit width, this setup resulted in a spectral resolution R of about 600.\\

To remove the sky lines, we used the chopping mode of ISAAC (with a chopping throw of 20" along the slit) combined with nodding. This resulted in a series of two consecutive frames A and B shifted by 20" each and composed of one positive and one negative sky-subtracted images. Subtracting one frame from the other produces a new frame with 2 negative images and one positive with twice the intensity of the negative ones.
The data were cleaned from cosmic-rays, flat-fielded, wavelength calibrated using a Xenon-Argon lamp and corrected for distortion along the wavelength axis. This was done using routines from the IRAF and MIDAS astronomical software packages.
 Finally, the data were corrected for telluric lines and calibrated in relative flux. This resulted in a number of spectra of the brightest sources of the region that we extracted to test the validity of our calibrations.\\

\begin{figure}%[h]
\includegraphics[width=24pc]{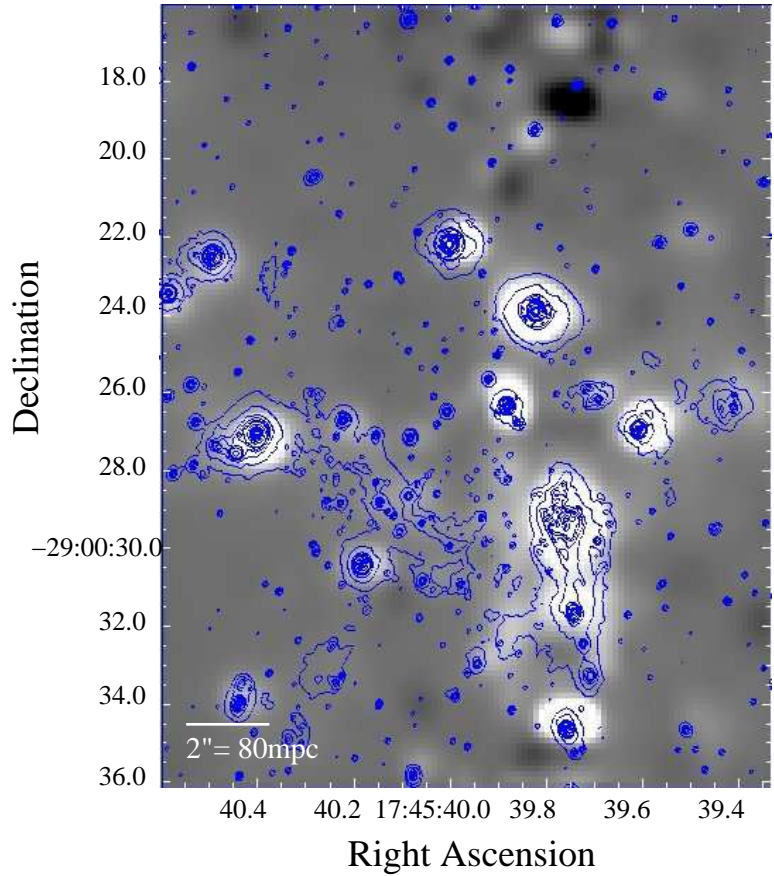}  %line60_960_contgcL_small.eps}
\caption{\label{contNACO} L-band emission image obtained by integrating the spectra of the data cube all over the wavelength range, with contours of a NAOS-CONICA L'-band image of the central parsec. The surface brightness in L'-band is given in contours from 0 to 2.87 Jy/beam in steps of 190 mJy/beam 
(one beam corresponds to one angular resolution element, i.e., FWHM 95 milli-arcsec).
}

\end{figure}

Since our aim was to build a data cube from a total of 26 slit positions, we used the data reduction parameters obtained for the individual spectra extracted from one frame in order to apply a mean value to the whole frame. This procedure was performed for each frame.\\
For each slit position, we shifted the reduced frames along the slit axis to match the positions of the trace image with each other. Then, we added up the shifted frames in order to increase the signal-to-noise. Using the DPUSER software\footnote{Developped by Thomas Ott http://www.mpe.mpg.de/~ott/dpuser/; see also Eckart \& Duhoux (1991).}, we created a data cube by shifting the final slit position frames with respect to each other in order to match the relative positions of the objects in the field. This final stage is done by eye to an accuracy of a fraction of an arcsecond. As a matter of fact, we show in Fig.\ref{contNACO} the integrated map along the entire wavelength range (after smoothing it with a Gaussian and a boxcar) over which, we overlaid contours of an  L'-band image of the half central parsec obtained with the NAOS-CONICA (NACO) camera located on the UT4 VLT/ESO telescope. The fit of the two images is very satisfying and shows that we managed to reproduce the field with an accuracy well below one arcsecond, which is the typical angular resolution of the data. For more details on the data reduction steps, see Moultaka et al. (2015).

\section{The line of sight extinction in the L-band}

\subsection{Derivation of the line of sight extinction spectrum}\label{derivation}

\begin{figure*}

\includegraphics[width=54pc]{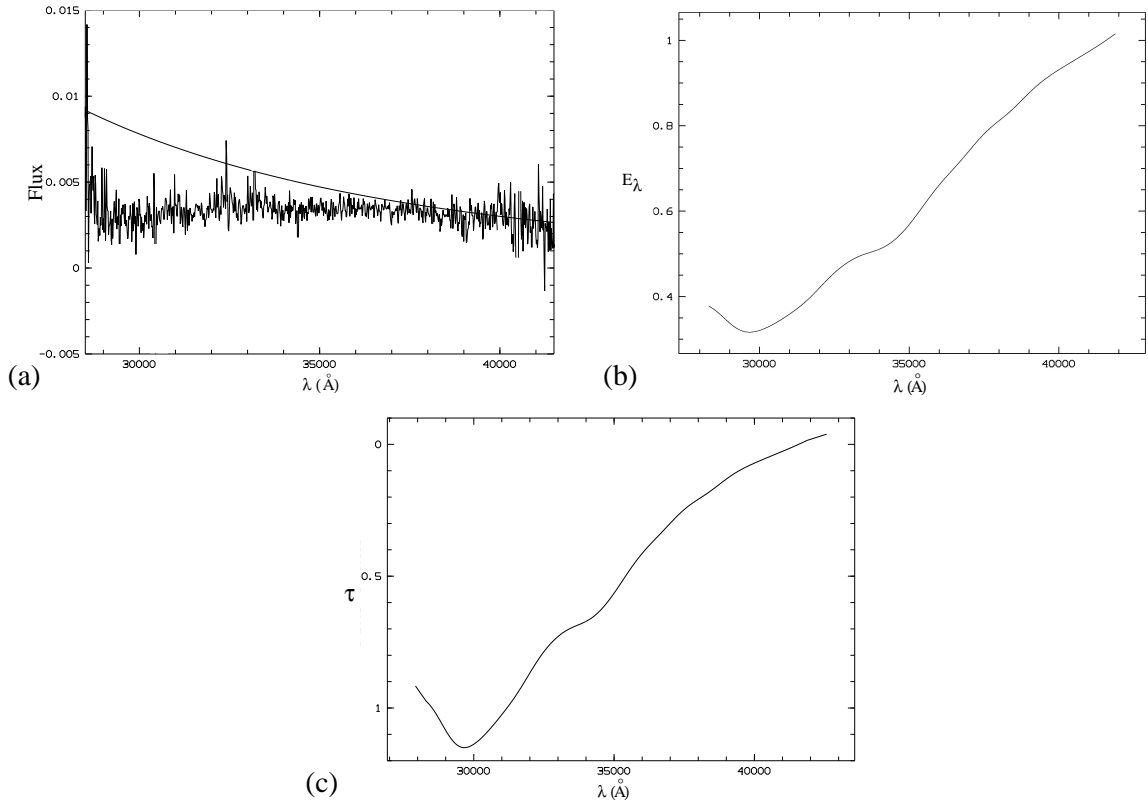}
\caption{\label{COstarspec} 
(a) L-band spectrum of the CO-star used to derive the calibrator of the foreground extinction. A blackbody continuum of 3600K temperature is overlaid. Note the absorption features at 3 and 3.4-3.48$\mu$m. - (b) Extinction spectrum obtained from the division of the observed CO-star spectrum by that of the blackbody. - (c) Optical depth spectrum derived from the extinction spectrum in (b).}

\end{figure*}

In a former paper (Moultaka et al. 2004), we derived a spectrum of the foreground extinction in the wavelength range of the SL filter of ISAAC. To that end, we used the L-band spectrum of a late type star showing CO bandheads in its K-band spectrum. 
In the following we will refer to this object as the "CO-star". Owing to the shape of its near-infrared spectrum and to the presence of this molecular feature, it can be safely assumed to have an effective temperature, of typically 3600K, corresponding to an M0-type star. Moreover, it is close to the studied region at an angular distance of 13" (i.e. $\sim 0.5$pc) from SgrA$^\star$, is located at the edge of SgrA$^\star$ West, outside the mini-spiral and the CND structures. Its L-band spectrum does not show excess emission. Hence, we can assume that it is not affected by local extinction and that its L-band spectrum, shown in Fig.~\ref{COstarspec}(a), is only extincted by the line of sight material. One can notice that the observed CO-star spectrum shows two absorption bands at $\sim 3 \mu$m and $\sim3.4-3.48 \mu$m due to the presence of water ice and hydrocarbons along the line of sight. \\

In order to derive the line of sight extinction spectrum, we created a blackbody spectrum of 3600K temperature and shifted it by a multiplicative constant to match that of the observed CO-star around 4.2$\mu m$. At this wavelength, we can safely assume that the continuum is free of absorption features and therefore matches the continuum of a non absorbed spectrum. Nevertheless, this assumption does not mean that the spectrum is not absorbed by a constant continuum extinction over the whole wavelength range (see explanation hereafter). 
We then, divided the observed spectrum by that of the shifted blackbody. Consequently, the resulting spectrum represents the wavelength-dependent extinction along the line of sight in the L-band domain that accounts only for the extinction due to the absorption features (see Fig.~\ref{COstarspec}(b)). Indeed, if we call $I_{{CO-star\, obs} \lambda}$, the flux at wavelength $\lambda$ of the observed spectrum of the CO-star and $I_{{CO-star\, intr} \lambda}$ its intrinsic flux at the same wavelength (that is ideally a blackbody emission of about 3600K temperature), then we have:
\begin{equation}
I_{{CO-star\, obs} \lambda} = I_{{CO-star\, intr} \lambda} * [E_{\lambda}] * k 
\label{I_obs} 
\end{equation} 
where $E_{\lambda}$ is the line of sight extinction spectrum and $k$ a constant accounting for an additional constant continuum due to dust extinction. Here $k$ is assumed to be equal to 1 since its value doesn't influence the results we present in the following sections.\\
The optical depth spectrum $\tau(\lambda)$ corresponding to the line of sight extinction is shown in  Fig.~8 of Moultaka et al. (2004) and in Fig.~\ref{COstarspec}(c). It is obtained via the definition: $\tau(\lambda)=-ln(\frac{I_{{CO-star\, obs} \lambda}}{I_{{CO-star\, intr} \lambda}})=-ln([E_{\lambda}]*k)$.

\subsection{Quantifying the foreground extinction in the L-band}

From the L-band line of sight extinction spectrum shown in Fig.~\ref{COstarspec}(c), we derive a mean optical depth of about 0.55 (or a mean extinction E$_L$ of 0.57). This corresponds to an extinction A$_L \sim 0.6$~mag. If we consider the extinction law by Martin \& Whittet~(1990) (obtained from observations of stars in the solar neighborhood and in the $\rho$ Oph cluster), the above value of A$_L$ will be consistent with an extinction in the K-band of A$_K \sim 1.38$~mag. On the other hand, it will be consistent with a K-band extinction of A$_K \sim 1.16$~mag if we use the extinction law of Rieke \& Lebofsky (1985) resulting from observations of stars in the Galactic Center. This is less than the mean value of about 2.46~mag obtained toward the Galactic Center by Sch\"odel et al.~(2010). We conclude that the absorption features do not account alone for the foreground extinction but an important contribution comes from the dust continuum absorption as well. This component is represented by the constant $k$ in equation~\ref{I_obs}. As a matter of fact, the continuum extinction by dust was not taken into account since we assumed a constant $k=1$ in section~\ref{derivation} when deriving the line of sight extinction spectrum. \\

Considering the A$_K$ value given by Sch\"odel et al.~(2010), we can estimate the value of the constant $k$. If we calculate the L-band extinction using the extinction law by Rieke \& Lebofsky, we obtain an A$_L \sim 1.27$mag. This is equivalent to an optical depth of $\tau_{L-Sch}=1.17$. Doing the same calculation but using Martin \& Whittet's law, we find an A$_L \sim 1.07$ mag implying an optical depth of $\tau_{L-Sch}=0.98$. In order to match the extinction A$_K$ of Sch\"odel et al.~(2010), the required constant, $\tau_{add}$, to be added to the optical depth spectrum of Fig.~\ref{COstarspec}(c), is then $\tau_{add}=\tau_{L-Sch}-0.55$. We obtain $\tau_{add}=0.62$ (for Rieke \& Lebofsky'law) and $\tau_{add}=0.43$ (for Martin \& Whittet's law). The corresponding constant $k$ of equation ~\ref{I_obs} is then derived through $k=e^{(-\tau_{add})}$. In the case of Rieke \& Lebofsky's law, we find $k=e^{(-0.62)}=0.54$, and in the case of Martin \& Whittet's law, we get $k=e^{(-0.43)}=0.65$. \\

\subsection{Validation of the line of sight extinction calibration}\label{validation}

In Moultaka et al. (2004), we obtained spectra of a dozen of bright sources located in the central parsec that we corrected for the foreground extinction. To that end, we divided the observed spectra by the calibrator extinction spectrum described previously and shown in Fig.~\ref{COstarspec}(b). The resulting spectra were then fitted by blackbody continua. The best fits were obtained with blackbody temperatures that agree very well with the known temperatures of the sources (or of the dust they are embedded in), obtained from NIR spectroscopy or imaging. \\
On the other hand, we also fitted the non-corrected spectra for the foreground extinction by blackbody continua reddened with a varying extinction $A_K$. The temperatures of the best fits matched very well those found previously using the calibrator extinction spectrum (see Tab.~\ref{Tabtemperatures}). Also, the mean value of the extinction $A_K$ obtained from the best fits agrees with the value of about 3 mag derived from imaging data (Sch\"odel et al. 2010) (see Tab.~\ref{Tabtemperatures}). 
These results show that the calibrator extinction spectrum is a very good approximation of the foreground extinction.

\begin{table}
\begin{center}
\label{Tabtemperatures}
\caption{Table summarizing the results of the two methods used in Moultaka et al. (2004) to estimate the foreground extinction.}
\begin{tabular}{rccr}
\tableline\tableline
Source & Temperature 1\tablenotemark{a} & Temperature 2\tablenotemark{b} & $A_{K}\tablenotemark{b}$ \\
\tableline
IRS 1W & 1200 &  900 & 3.30\\
IRS 3  & 1200 &  800 & 4.20\\
IRS 7  & 1100 & 2100 & 3.55\\
IRS 9  & 2000 & 1900 & 3.10\\
IRS 13  &  5000 & 1300 & 3.00\\
IRS 13N  & 1500 & 1000 & 3.90\\
IRS 16C  & 26000 & 22600 & 3.45\\
IRS 16CC  & 20000 & 20200 & 4.00\\
IRS 16NE  & 20000 & 22400 & 3.25\\
IRS 16SW  & 20000 & 22500 & 3.25\\
IRS 21  & 1000 & 1200 & 3.95\\
IRS 29  & 1200 & 1600 & 3.70\\
\tableline
\end{tabular}

\tablenotetext{a}{The temperature is obtained from the best fit of a blackbody spectrum to the corrected spectra for the foreground extinction using the calibrator extinction spectrum.}
\tablenotetext{b}{The temperature and K-band extinction A$_K$ are obtained from the best fit of a reddened blackbody spectrum to the observed source spectra.}

\end{center}
\end{table}

In the present work, we have an L-band data cube that covers the entire field of the central half parsec. Using the calibrator extinction spectrum, we built a data cube corrected for the foreground extinction. To this end, we divided all the spectra of the field at each spatial pixel (or equivalently, all the lines in the frames corresponding to the different slit positions) by the extinction spectrum. We show in Figs.~\ref{Lbandcorr} and~\ref{Lband} the maps of the integrated fluxes along the wavelength range obtained from the corrected and non-corrected data cubes for extinction, respectively. The similarity of both maps shows that there is no over-correction of the extinction in the observed field.

\begin{figure*}%[h]
\includegraphics[width=40pc]{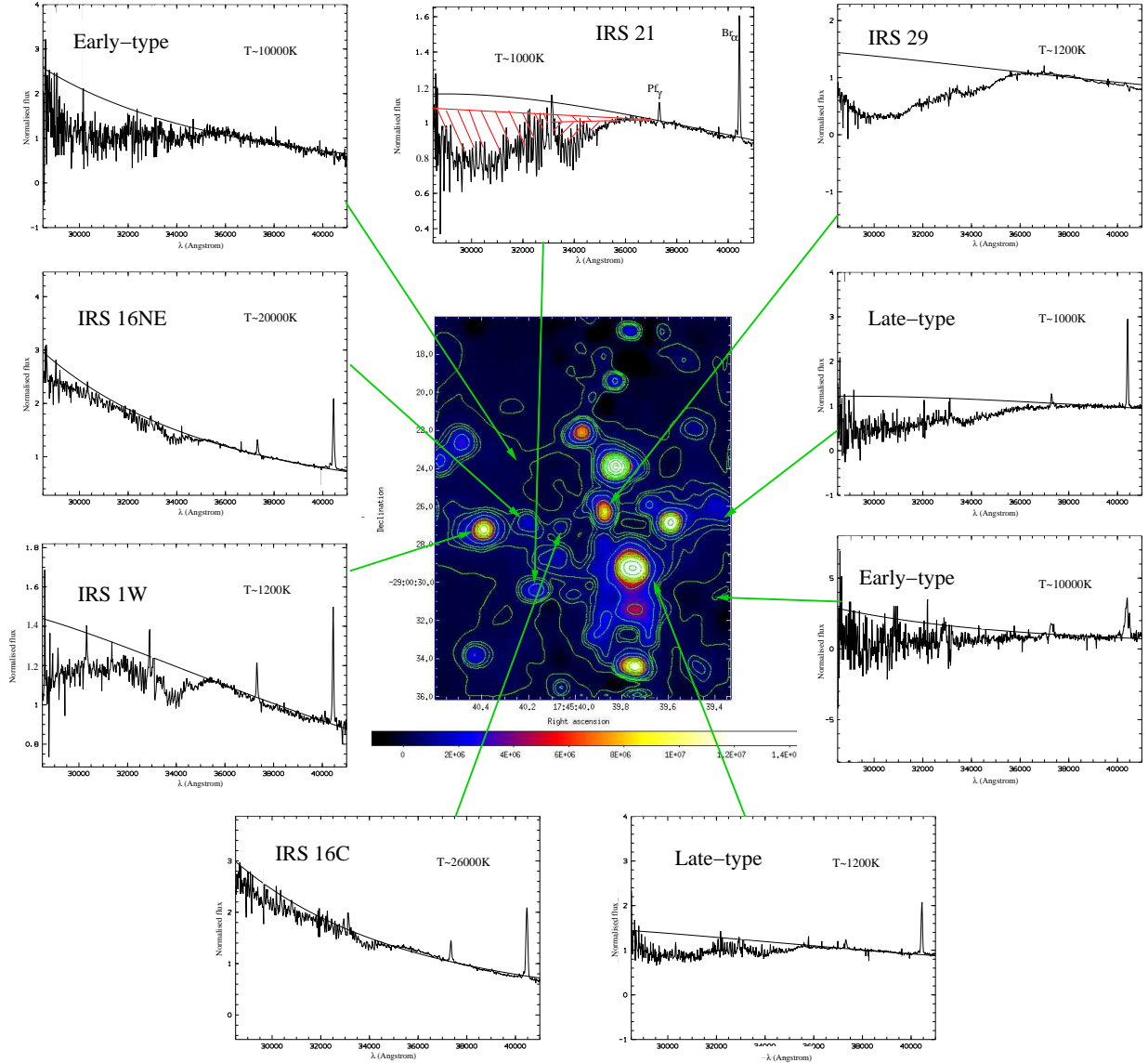} 
\caption{\label{newfig} L-band spectra corrected for the foreground extinction of the bright sources IRS1W, IRS 16C, IRS16NE, IRS21, IRS29 and late-type and early-types stars referenced by Buchholz et al. (2009). The corrected spectra are well fitted by the continuum emission of a blackbody. The temperature of the best fitted blackbody continua are in agreement with the nature of the sources. On the spectrum of IRS~21, we show in red, the adopted continua that were used to derive the optical depths of the 3.0 micron and the 3.4-3.48 micron features. The dashed red area indicates the corresponding absorptions. We also indicate the Br$_\alpha$ and Pf$_\gamma$ emission lines. The typical relative uncertainty on the spectral data points in the blue part is of 25\% and in the red part about 10\%. }

\end{figure*}

\begin{figure}

\includegraphics[width=24pc]{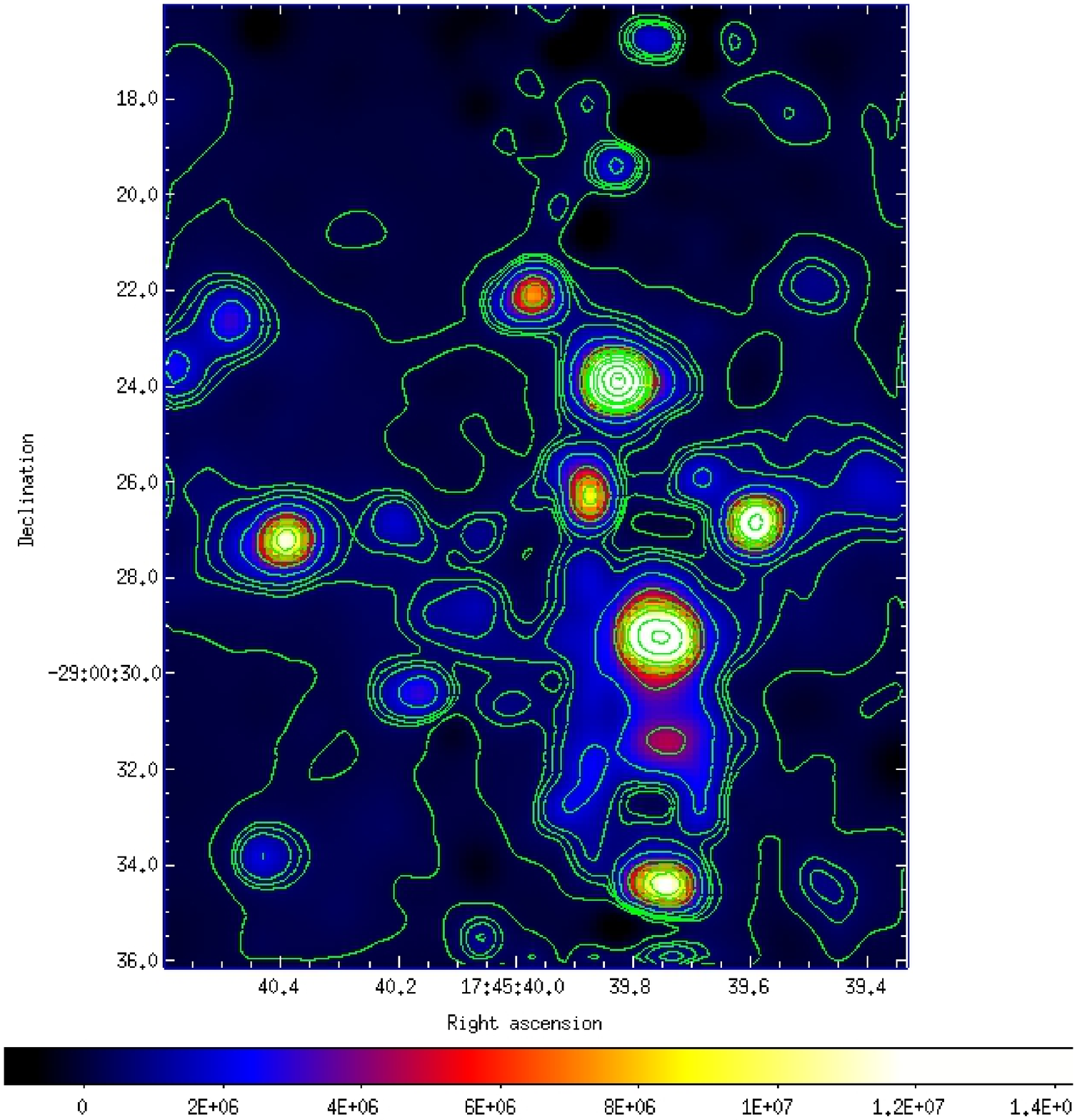}
\caption{\label{Lbandcorr} Integrated emission along the L-band spectral domain obtained from our L-band data cube corrected for the foreground extinction. Contours are also shown for clarity. The surface brightness in L-band is given in contours from 0 to  4.54 Jy/beam in steps of 324 mJy/beam (one beam corresponds to one angular resolution element, i.e., FWHM 0.7 arcsec).
}
\end{figure}

To test the validity of the calibrator extinction spectrum in the region, we fitted blackbody continua to a number of spectra of early and late-type stars located all around the central half parsec and outside the mini-spiral. These stars have been classified as early-type or late-type by Buchholz et al. (2009) by means of near-infrared photometry and spectral energy distributions. The best fits were obtained with temperatures of about 10~000K to 40000K for the early-type stars and about 1000K to 5000K for the late-type stars, which is in agreement with the temperature of their spectral types. This shows that our calibrator extinction spectrum of the line of sight extinction is also valid over the entire region. In Fig.~\ref{newfig} we show a number of spectra for each of the stellar types presented by Buchholz et al. (2009) with the best blackbody continuum fits overlaid. We also show the spectra of the bright infrared sources IRS 1W, 16C, 16NE, 21 and 29.\\
On one side, this result validates our calibration of the line of sight extinction spectrum and on the other side, our L-band spectra confirm the nature of the observed stars. Our results also agree with Scoville et al. (2003) and Sch\"odel et al (2010) who found a rather smooth distribution of the extinction toward the Galactic Center on 1" to 2" scale in the near infared, with a variation of A$_K$ not exceeding 0.5 mag.

\section{Data Cube analysis}

In the previous section, we validated our calibration of the foreground extinction in the L-band. 
Thus, we can safely use the corrected data cube for the line of sight extinction derived in section~\ref{validation}. 
In the following, we present and analyze different maps obtained from the corrected data cube and compare them to their 
counterparts obtained from the non-corrected data cube.

\subsection{Water ice and Hydrocarbon maps}\label{maps}

We aim at deriving the distribution of the local water ice and hydrocarbon features in the half parsec around SgrA$^\star$. 
We assume that the continua of the hydrocarbon and water ice absorptions can both be approximated by straight lines (see Fig.\ref{newfig}). 
These assumptions are justified by the following facts: 
\\
1) The hydrocarbon absorption is located in the red wing of the water ice absorption and the latter can be approximated by a straight line from 3.32$\mu$m to 3.77$\mu$m. This is observed, for example, in the GC spectra by Moultaka et al. (2004) (see figures 2, 3, 4, 9, 10 and 11 of that paper), in figure 7 by Chiar et al. (2002) (showing a mean GC ice feature as well as three water ice models) and in the spectra of young stellar objects and quiescent molecular clouds with background stars, shown in figure 3 by Noble et al. (2013). 
\\
2) The relative uncertainty on the optical depths of the water ice feature is less than $10\%$, if we assume a straight line continuum from 2.84$\mu$m to 3.77$\mu$m instead of a blackbody continuum or instead of fitting a laboratory water ice spectrum. Indeed, this deviation can be estimated from the figures mentioned above. 
Moreover, in the Taurus molecular cloud complex Murakawa et al. (2000) estimated that 
due to this approximation of the continuum, the relative uncertainty
in deriving the optical depths of the water ice absorption is about  $5\%$. 
Note, finally, that the wavelength interval of the hydrocarbon absorptions includes the 3.53 micron methanol 
ice absorption but this feature is not clearly detected in our spectra. The optical depths are calculated through the definition 
\begin{equation}
\label{tau}
\tau = -ln(\frac{F_{obs}}{F_{cont}})
\end{equation}
 where $F_{obs}$ is the integrated observed flux over the absorption feature and $F_{cont}$ is the integrated continuum flux along the spectral feature. The map of the optical depth of the water ice absorption is obtained after subtracting that of the hydrocarbon absorption since 
the latter is located in the wing of the water ice feature (see Fig.\ref{newfig}).\\
 The advantage of using optical depths is that they are independent of the continuum intensity. Their values can, therefore, be directly compared, regardless of the background brightness.\\

\begin{figure}%[h]

\includegraphics[width=24pc]{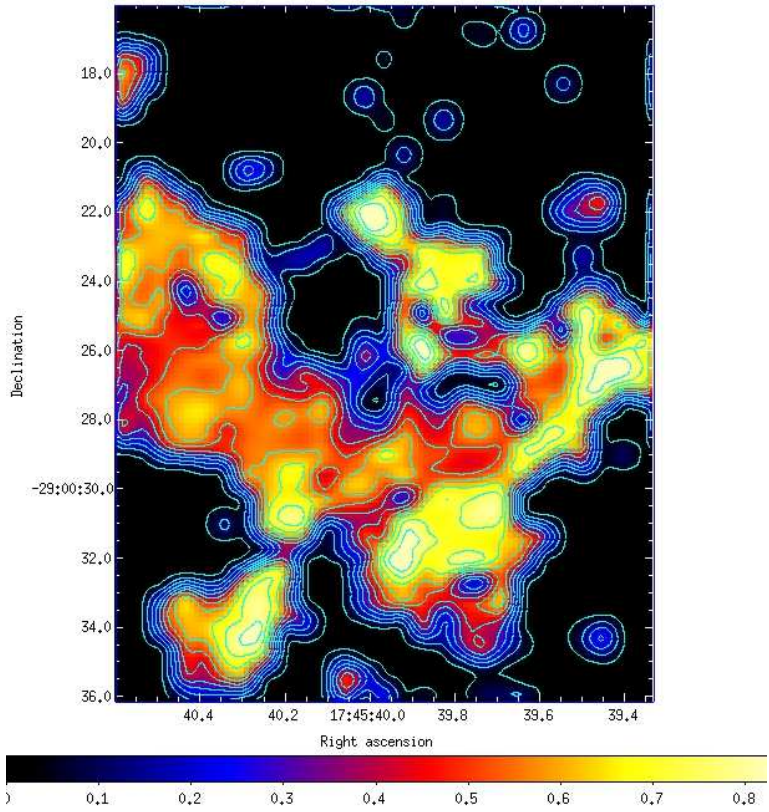}
\caption{\label{tauIcecorr} Optical depth map of the water ice absorption feature extracted from our extinction corrected data cube. Contours are also shown for clarity. Contour levels are $\tau$ = 0.03, 0.88 in steps of 0.08}.

\end{figure}

\begin{figure}

\includegraphics[width=24pc]{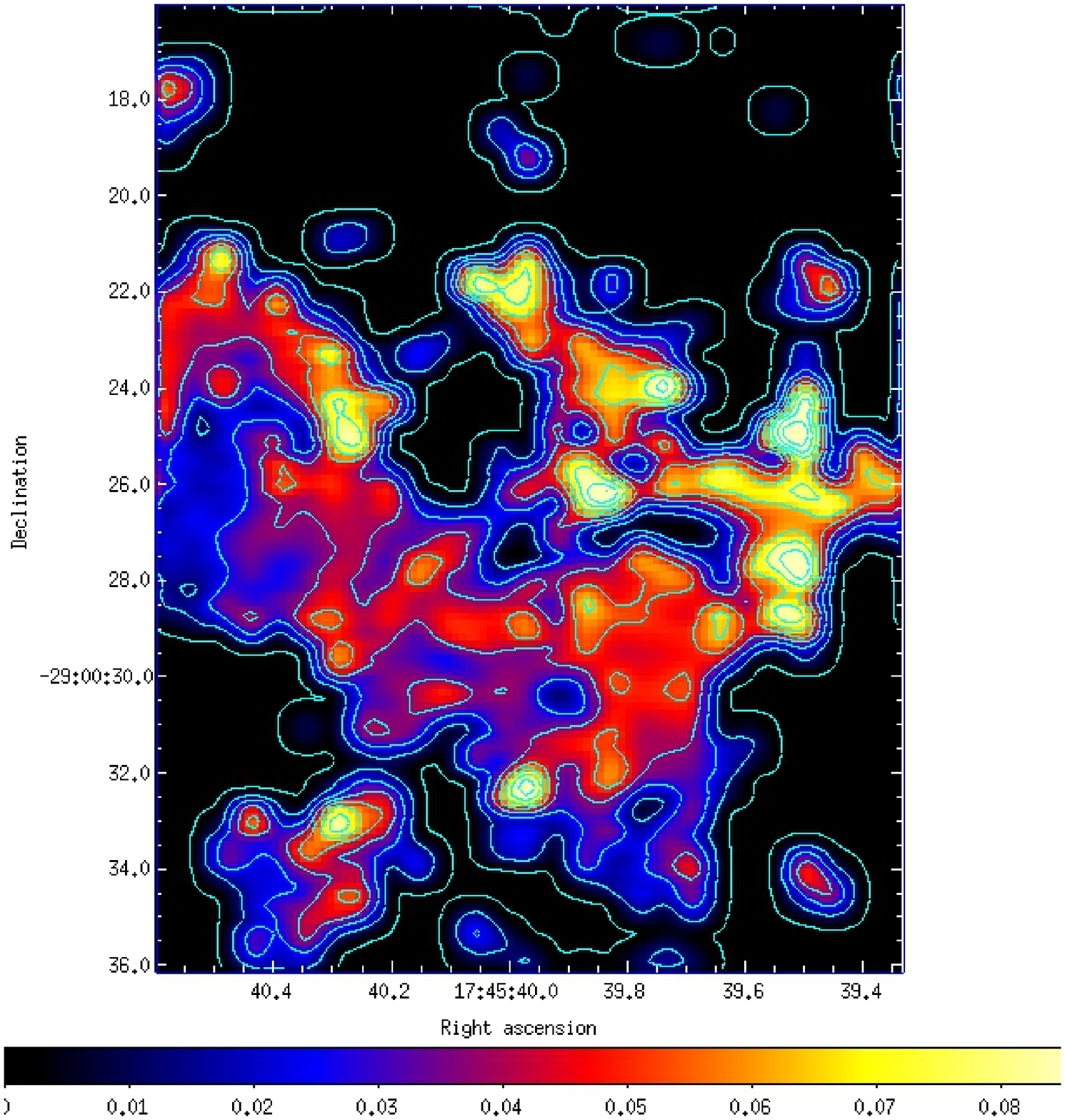}
\caption{\label{tauHydcorr} Optical depth map of the hydrocarbon absorption feature at 3.4-3.48 $\mu$m extracted from our extinction corrected data cube. Contours are also shown for clarity. Contour levels are $\tau$ = 0.001, 0.01 to 0.11 in steps of 0.012.}

\end{figure}

The resulting maps obtained from the data cube corrected for the foreground extinction are shown in Figs.~\ref{tauIcecorr} and~\ref{tauHydcorr} for the water ice and the hydrocarbon features, respectively. 
In Figs.~\ref{tauIce} and~\ref{tauHyd} (Appendix A), we show the {optical depth maps of the same absorption features} obtained from the non-corrected data cube for extinction. The water ice maps corrected and non-corrected for extinction are very similar as can be seen 
by comparing Figs.\ref{tauIcecorr} and~\ref{tauIce}. Also, we find similar distributions of the 3.4-3.48$\mu m$ bands in the corrected and non-corrected maps (see Figs.~\ref{tauHyd} and \ref{tauHydcorr}). This suggests that the line of sight extinction is not over-estimated and the data cube is not over-corrected since no negative values are encountered in the maps. \\

The first striking result is the presence of residual absorptions in the corrected maps.
Since we corrected for the foreground effects this is indicative of absorbing material possibly in the local medium of the Galactic Center.

\begin{figure}%[h]
\includegraphics[width=24pc]{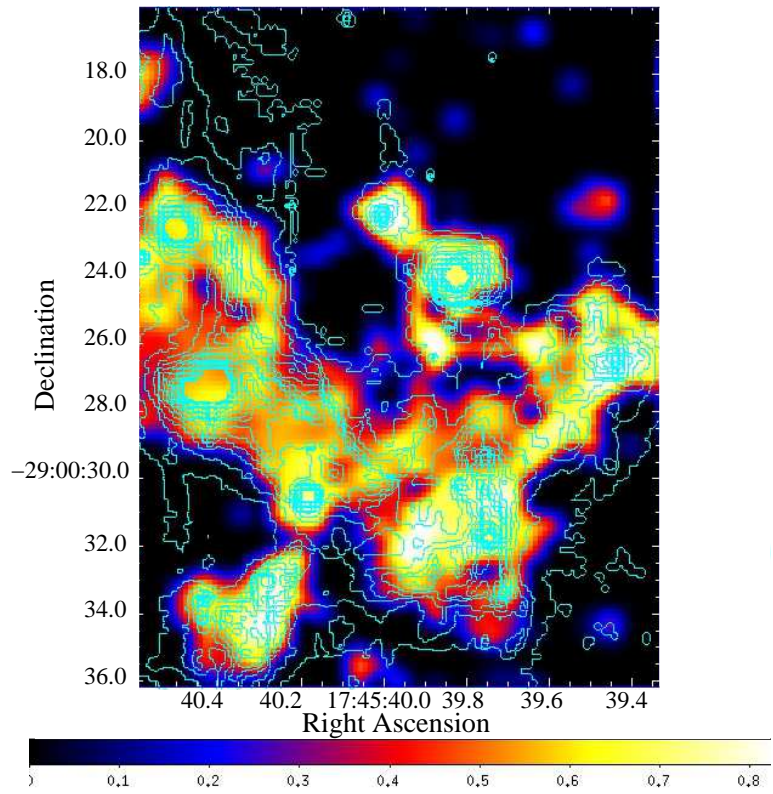} 
\caption{\label{visir} Optical depth map of the water ice absorption with contours of the VISIR 8.6$\mu m$ image from Viehmann et al.~(2006). The surface brightness at 8.6 $\mu$m is given in contours from 0 to 253 Jy/beam in steps of 13.3 mJy/beam (one beam corresponds to one angular resolution element, i.e., FWHM 0.8 arcsec).
}
\end{figure}

\begin{figure}%[h]
\begin{minipage}{20pc}
\includegraphics[width=20pc]{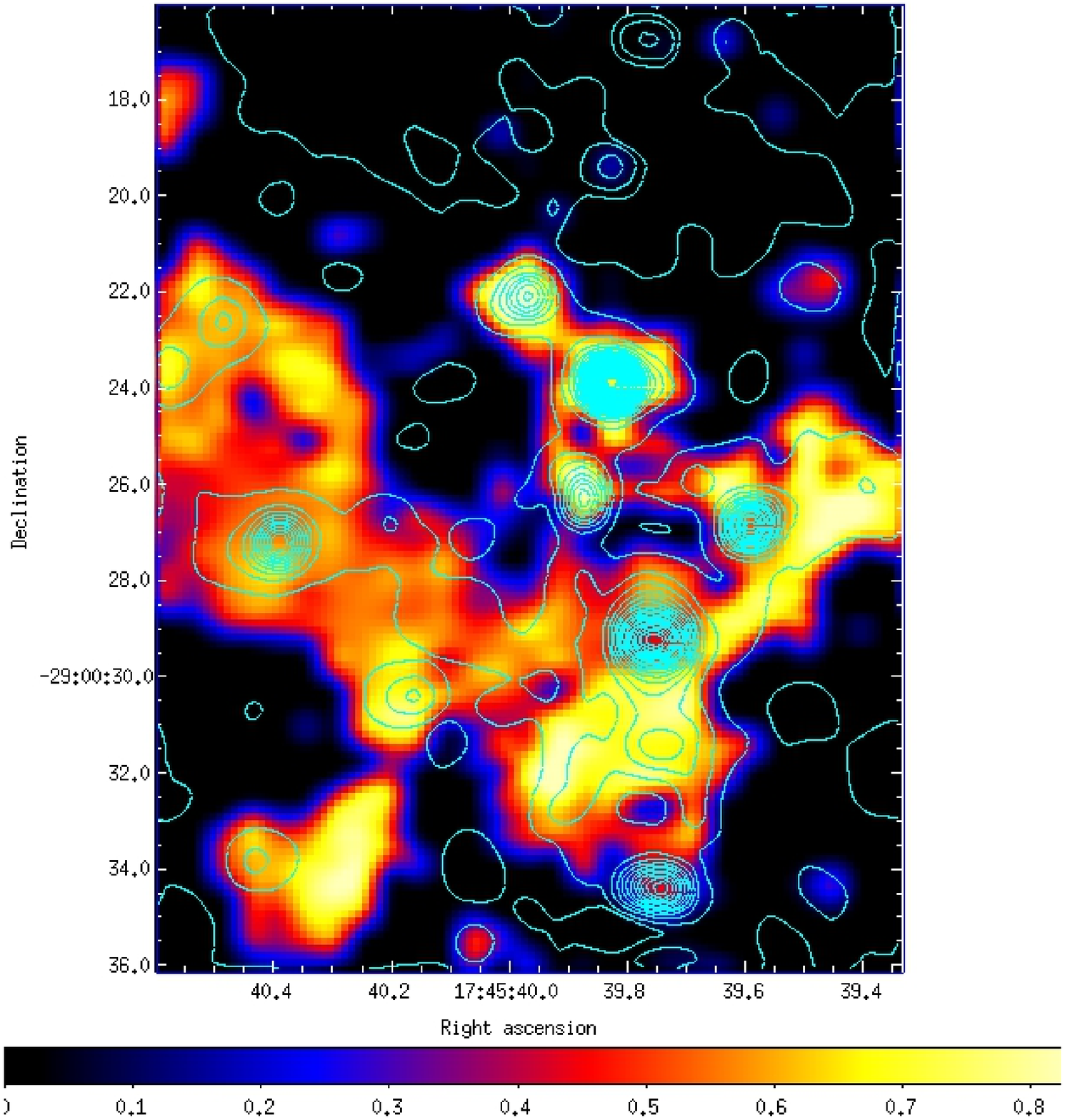} 
\caption{\label{Icecontline} Optical depth map corrected for foreground extinction of the water ice absorption feature with contours of the integrated L-band emission. The surface brightness in L-band is given in contours from 0 to  4.54 Jy/beam in steps of 324 mJy/beam (one beam corresponds to one angular resolution element, i.e., FWHM 0.7 arcsec).}
\end{minipage}\hspace{2pc}%
\begin{minipage}{20pc}
\includegraphics[width=20pc]{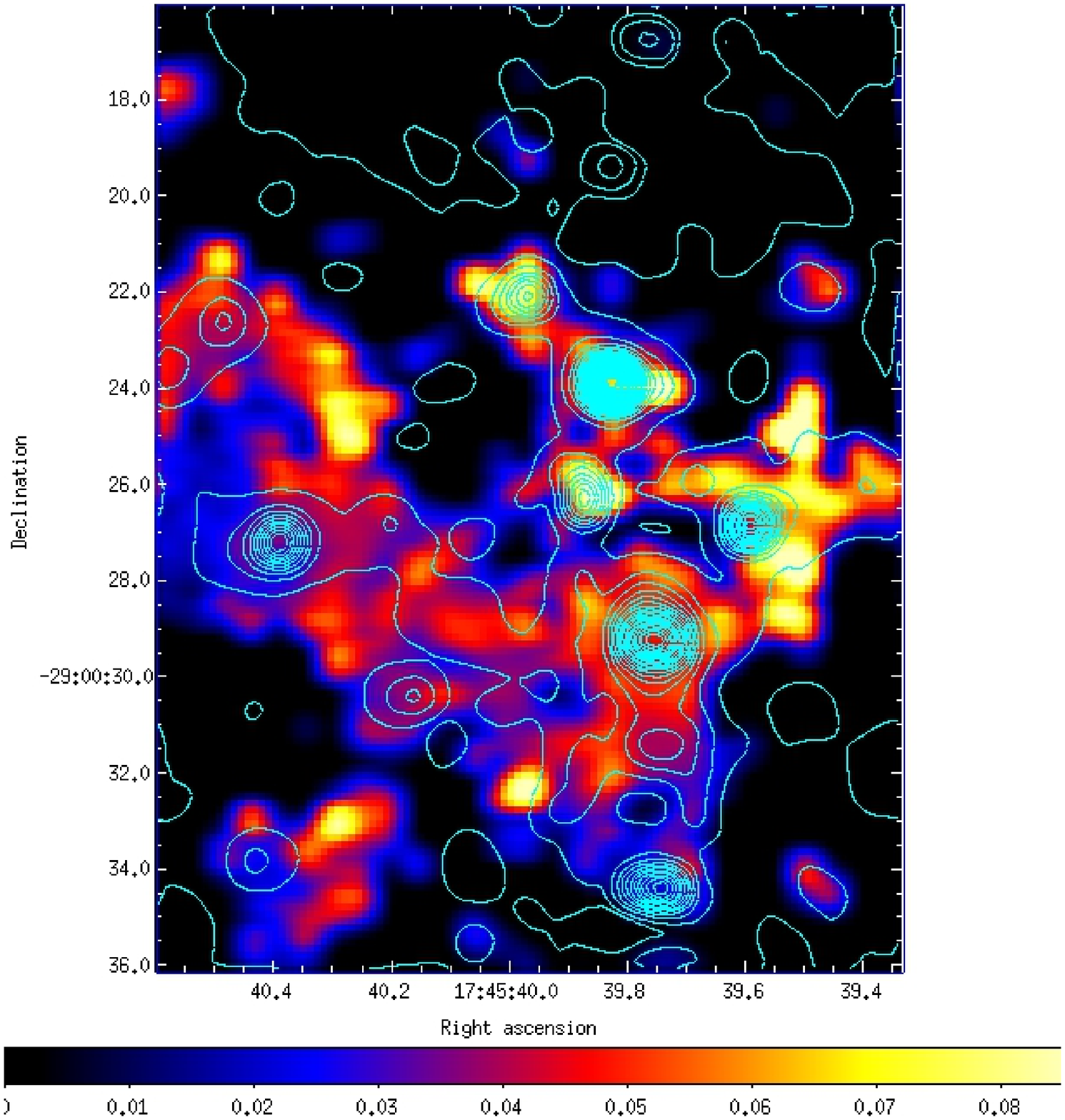}
\caption{\label{Hydcontline} Optical depth map corrected for foreground extinction of the hydrocarbon absorption feature with contours of the integrated L-band emission. The surface brightness in L-band is given in contours from 0 to 4.54 Jy/beam in steps of 324 mJy/beam (one beam corresponds to one angular resolution element, i.e., FWHM 0.7 arcsec).}
\end{minipage} 
\end{figure}

On the other hand, the corrected maps show that the 3 and 3.4-3.48 micron features obviously trace the mini-spiral structure. 
This is also highlighted in Fig.~\ref{visir} where we overlay the water ice corrected map with 
contours of the infrared image at 8.6$\mu m$ obtained with VISIR/ESO imager (Viehmann~et~al.~2006).
Since emission at 8.6 micron is dominated by dust, figure~\ref{visir} shows that water ices trace the dust very well except in two regions. As a matter of fact, in the north-eastern part of the field, the 8.6 micron emission is bright while water ices are not detected. In the north-western part we find the opposite behaviour. The mismatch in the first case is either due to the absence of water ices stuck on the dust grains or, more probably, to the absence of bright background sources resulting in a low Signal-to-Noise ratio of the spectra. In the north-western part, we detect the ice absorption since there are bright background sources (see Fig.\ref{Icecontline}) but the 8.6$\mu m$ emission is very faint or absent. This is probably due to the small amount of dust in the region which is consistent with the small optical depths found in our map at the same positions.

In addition, we show in Fig.~\ref{muzic} our water ice absorption map ({\it right}) in comparison with the high-pass filtered L'-band (at $3.8\mu m$) image from Mu$\check{z}$i\'c~et~al.~(2007) ({\it left}). The latter image highlights the compact components of the L'-band image obtained with NACO (NAOS/CONICA at ESO/VLT telescope) and reveals a large number of thin filaments in the mini-spiral. The authors interpret these features as the result of the interaction of wind from SgrA$^\star$ with dust in the mini-spiral. Here our map has a lower spatial resolution (about 1 arcsec) than that of Mu$\check{z}$i\'c~et~al.~(2007) which is of 100~mas. Despite this discrepancy, it is clear from Fig.~\ref{muzic}, that the peaks of the optical depth values of the water ice match very well the positions of the thin filaments of dust. Hence, water ice seems to trace the high density pockets of the dust. \\
On the other hand, the distribution of the water ice and the hydrocarbon absorptions also trace the dusty sources like IRS3, IRS7 and the north of IRS13 complex (see Figs.~\ref{Icecontline} and~\ref{Hydcontline}). Thus, these features are clearly associated with the dust components in the region. \\ 
All these results provide an additional validation of our calibration of the foreground extinction in the L-band. Indeed, it is unlikely that concentrations of dust along the line of sight happen to be located at the same projected distances to SgrA* as the local dust concentrations
in the Galactic Center region.\\  

Furthermore, our resulting maps suggest that the water ice and the hydrocarbon features can be associated with the neighboring infrared sources. Indeed, in Figs.~\ref{Icecontline} and~\ref{Hydcontline}, we overlay contours of the integrated L-band emission map on the optical depth maps; these figures show that peaks of absorptions are almost systematically offset by a few arcseconds (equivalent to a few angular resolution elements) from the bright sources of the L-band continuum map, except, maybe, for IRS~7 and in the southern part of the field. We interpret these observations as being a structure of mass loosing stars resulting from the interaction with the local ISM. 
This interaction may be enhancing their surrounding dust emission possibly by piling up the local dust. 
In the few cases where the peaks of the optical depth maps overlap those of the integrated emission map, the ISM is probably squeezed in the direction of the line of sight. \\

\begin{figure}%[h]
\begin{minipage}{20pc}
\includegraphics[width=20pc]{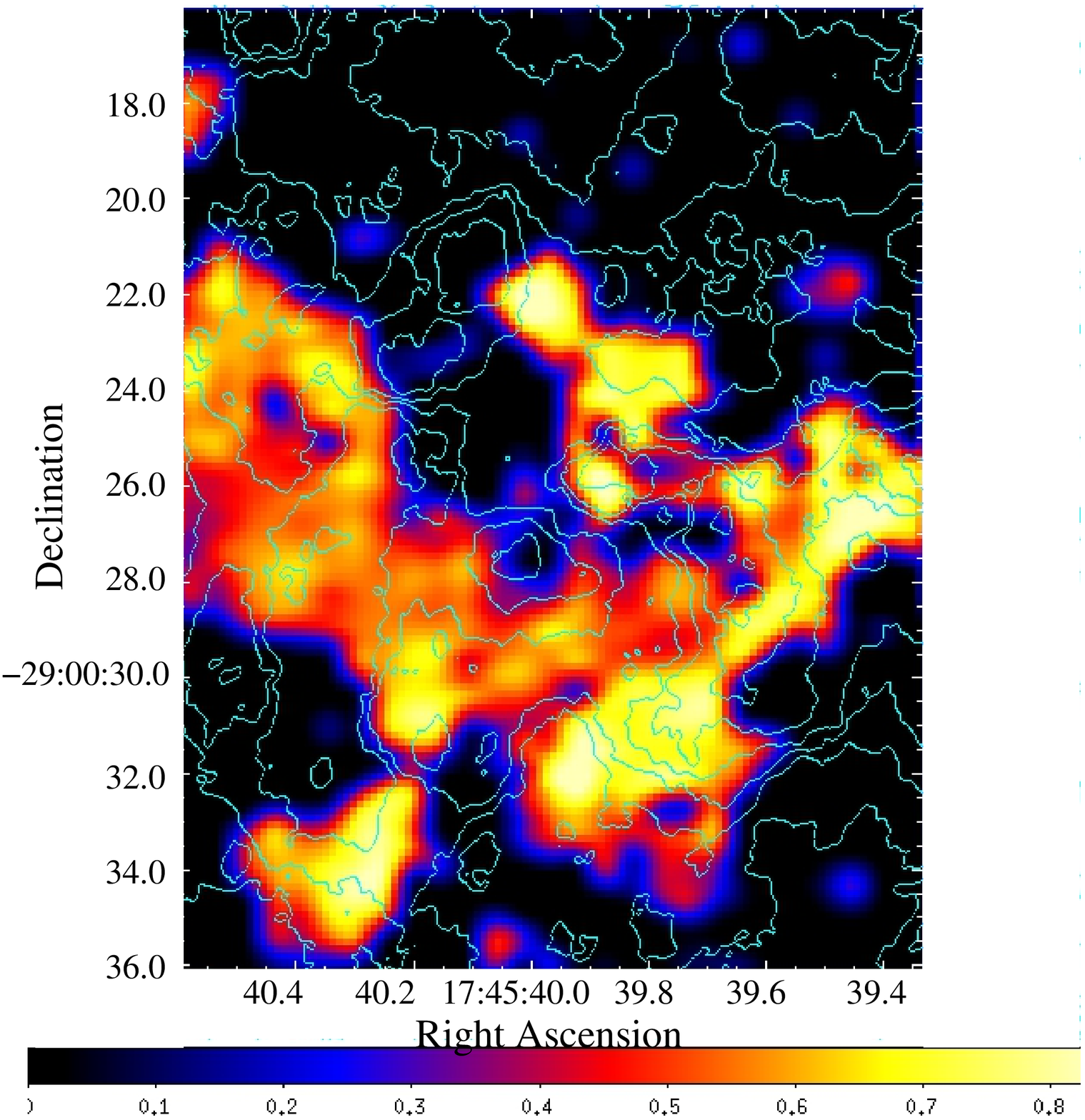}  
\caption{\label{IceKband} Optical depth map of the water ice absorption with contours of the K-band extinction image from Sch\"odel et al. (2010) overlaid. Contour levels are A$_K$= 0 to 3.26 in steps of 0.11.}

\end{minipage}\hspace{2pc}
\begin{minipage}{20pc}
\includegraphics[width=20pc]{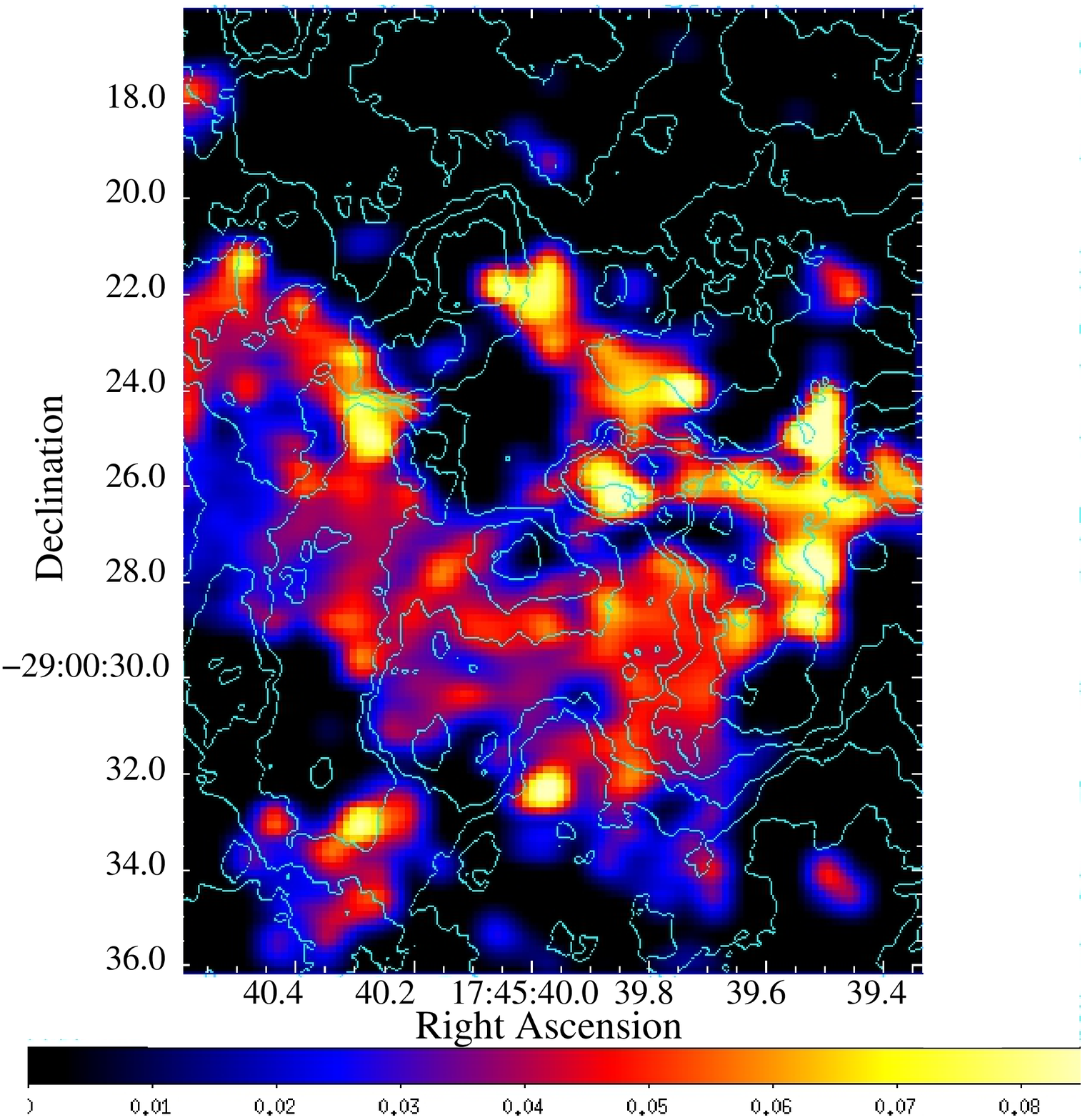}  
\caption{\label{HydKband}Optical depth map of the hydrocarbon absorption with contours of the K-band extinction image from Sch\"odel et al. (2010) overlaid. Contour levels are A$_K$= 0 to 3.26 in steps of 0.11.} 

\end{minipage} 
\end{figure}

\begin{figure*}%[h!]
\includegraphics[width=35pc]{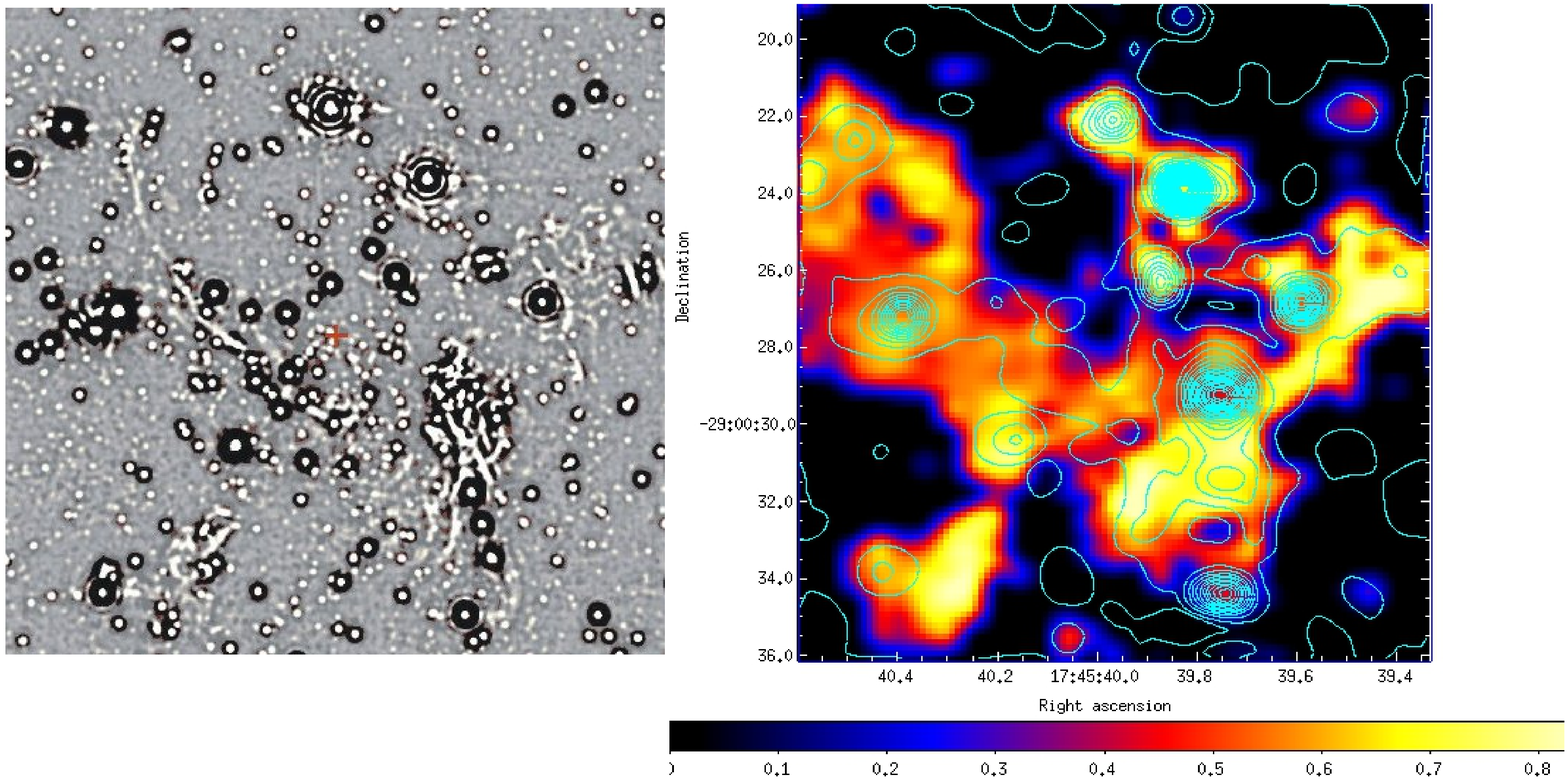}
\caption{\label{muzic} {\it (Left:)} For immediate and easy comparison with our new results shown on the right we present here the high-pass filtered L'-band image of the central half parsec obtained by Mu$\check{\bf z}$i\'c~et~al.~(2007). {\it (Right:)} Our water ice optical depth map at the same scale but with a lower spatial resolution. Contours of the integrated L-band emission are also shown with the same contour levels as in Fig.~\ref{Lbandcorr}. Colors are the same as in Fig.~\ref {tauIcecorr}.}  
\end{figure*}

We show in Figs.~\ref{IceKband} and~\ref{HydKband} the K-band extinction map derived by Sch\"odel et al. (2010) overlaid on our optical depth maps. These figures witness a full absence of correlation between the K-band map and our L-band extinction maps.
This result is a necessary condition if we claim that the absorption features occur in the local environment. Indeed, the extinction as traced by the K-band images is mainly due to continuum absorption due to the presence of dust along the line of sight. If the absorption features in the L-band were correlated with the K-band continuum absorption, this would imply that these absorptions are directly linked to the dust all along the line of sight. The fact that there is no correlation implies that, either, not all dust contains water ices/hydrocarbons or that the absorption features are local to the Galactic center and therefore diluted in the much higher global extinction due to the foreground and the local dust.

\subsection{Quantifying the local extinction}\label{quantLocal}

\begin{figure}%[h]
\begin{minipage}{24pc}
\includegraphics[width=24pc]{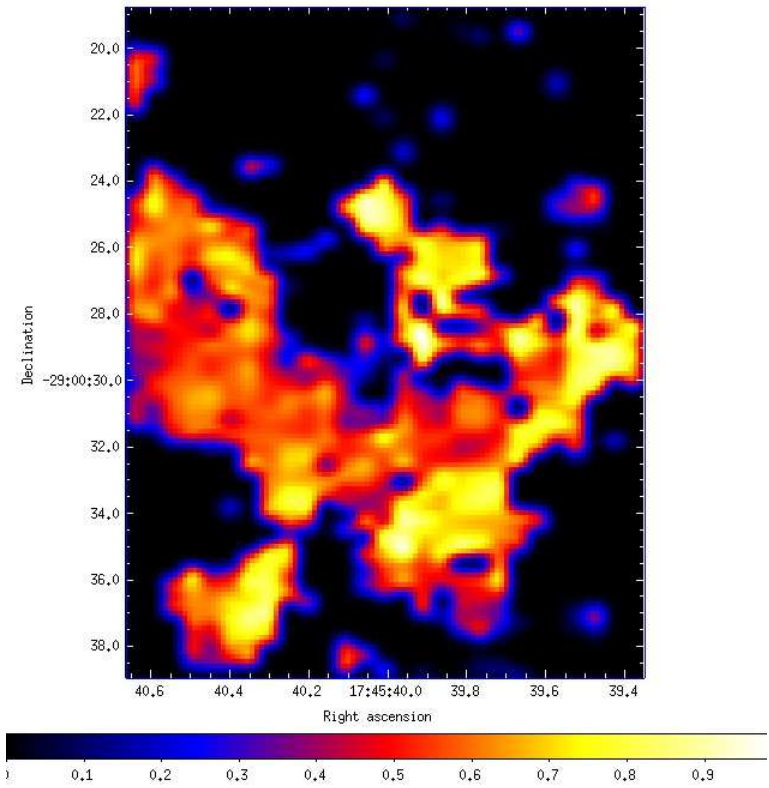}  
\caption{\label{tauLband} Smoothed version of the optical depth map of the extinction along the entire L-band corrected for the foreground extinction.}
\end{minipage} 
\end{figure}

In order to quantify the local extinction along the entire spectrum, we created the optical depth map of all the material absorbing over the full wavelength range. This map cannot be derived from the previous 3 and 3.4-3.48 micron maps because there is no correlation between the individual absorptions and the overall extinction of the spectrum. To this end, we calculated the optical depths from the corrected data cube for the foreground extinction, as in equ.~\ref{tau}; but in this case, the observed flux, $F_{obs}$, and the continuum flux, $F_{cont}$, are the observed and continuum/intrinsic fluxes integrated over the full wavelength range, respectively. The smoothed version of this map is shown in Fig.~\ref{tauLband}. We find that the optical depths vary from 0 to 1 in the whole region. This means that A$_L$ varies spatially from 0 to 1.08~mag corresponding to a variation of A$_K$ from 0 to 2.1~mag (for the case of the Rieke \& Lebofsky extinction law) and from 0 to 2.5~mag (for the case Martin \& Whittet extinction law). This is much higher than the 0.6-0.7~mag variation along the line of sight derived by Sch\"odel et al.~(2010) implying that probably most - but at least a significant part - of the residual absorbing material we observe in the L-band is located in the local medium of the central parsec.\\

\noindent{\bf Conclusion to sections \ref{maps} and \ref{quantLocal}}

The presence of water ices that we find in the local medium of the Galactic Center implies temperatures of the order of 10 to 80K (Chiar et al. 2002). This is consistent with our finding in Moultaka et al. (2009), and Moultaka et al. (2015), that CO ices are also probably located in the region as well.  
We showed in Moultaka et al. (2009) that, given the complexity of the region with the presence of bow-shock sources (e.g. Tanner et al. 2002, 2005), narrow dust filaments (e.g. Mu$\check{z}$i\'c et al. 2007, 2010) and dust embedded YSOs (Eckart et al. 2004, 2013), one cannot discard the presence of high-density pockets and high-optical depths where material at very low temperatures can also be present. These compact dusty structures can persist and survive while traveling through the central parsec since we showed that the travel time is shorter than the evaporation time of molecular clumps and disks in similarly harsh environments.\\

\subsection{Correlation between the dense and diffuse ISM}
The hydrocarbon and water ice corrected optical depth maps of Figs.~\ref{tauIcecorr} and~\ref{tauHydcorr} show a certain resemblance suggesting that these features are similarly distributed in the field. This idea was also invoked in our previous papers (Moultaka et al. 2004, 2005). In Fig.~\ref{figcorr}, we plot contours of the hydrocarbon absorption map over that of the water ice to highlight this trend. In order to quantify the correlation between the two absorptions, we derived the optical depths at each pixel of the maps corrected for foreground extinction, where the signal-to-noise ratio in the integrated L-band corrected map is larger than 10. Then we calculated the mean optical depth of the hydrocarbon absorption of all pixels for which the water ice optical depth is in {bins of 0.1 width. \\
We plot in Fig.~\ref{plotcorr}, the mean optical depths of the hydrocarbon feature as a function of the water ice optical 
depths in bin-widths of 0.1.
With a correlation coefficient of 0.85, this figure reveals a clear correlation 
between the water ice and the hydrocarbon optical depths. The best linear fit is shown in blue in Fig.~\ref{plotcorr}. This correlation suggests that the ISM presents itself as a - possibly clumpy - mixture of a dense and a diffuse medium.\\

\begin{figure}%[h]
\begin{minipage}{24pc}
\includegraphics[width=24pc]{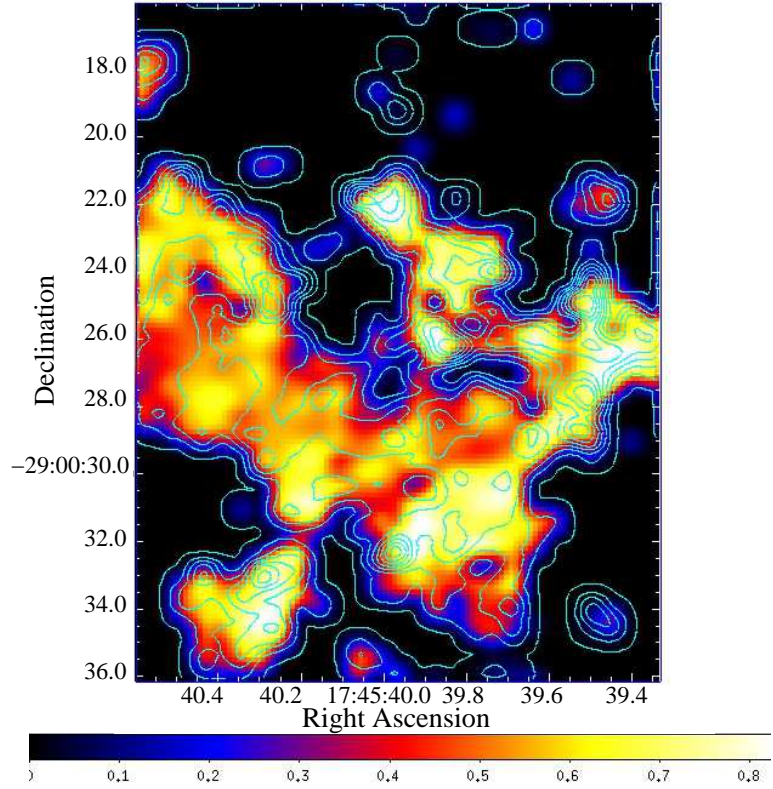} 
\caption{\label{figcorr} Optical depth map corrected for the foreground extinction of the water ice absorption with contours of the hydrocarbon absorption corrected optical depth map. Contour levels are $\tau$ = 0.001, 0.01  to 0.11 in steps of 0.012.}

\end{minipage}\hspace{6pc}
\end{figure}

\begin{figure}%[h]

\includegraphics[width=20pc]{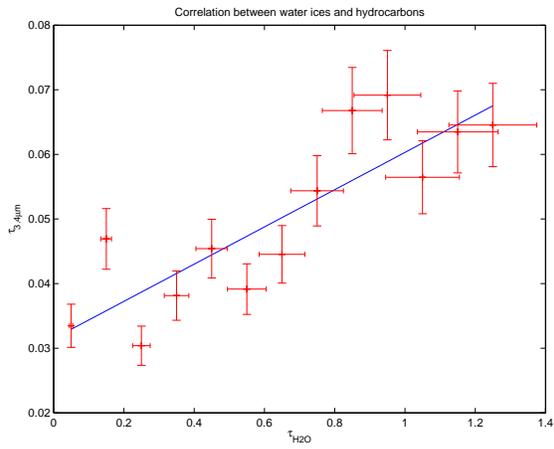} 
\caption{\label{plotcorr} Optical depth of the hydrocarbon versus the ice absorption features. This figure plots the mean intensities of all pixels of the hydrocarbon corrected map for which the water ice optical depths are given in bins of 0.1 width. The pixels considered in the maps are those at which the S/N ratio in the integrated L-band map, is higher than 10. The best linear fit is also shown in blue.}

\end{figure}

\subsection{Hydrogen emission line maps}
From the data cubes described previously, we also derive maps of the emission line strength for the Pf$_\gamma$ and Br$_\alpha$ lines located at 3.729$\mu m$ and 4.051$\mu m$, respectively (see an example of a spectrum in Fig.~\ref{newfig} and in Figure~13 of Moultaka et al.~2004). 
We define the line strength via $\frac{\int_\lambda I_{line}(\lambda)-I_{cont}(\lambda)}{\int_\lambda I_{cont}(\lambda)}$, where the integral along the emission line is calculated from 3.722$\mu m$ to 3.744$\mu m$ for the Pf$_\gamma$ line and from 4.028$\mu m$ to 4.061$\mu m$ for Br$_\alpha$); $I_{line}(\lambda)$ is the intensity of the line at each wavelength and $I_{cont}(\lambda)$ the intensity of the assumed continuum at the same wavelength. The continuum is approximated by a straight line connecting the mean continuum fluxes on the red and the blue sides of the line. They are calculated over 16 pixels (i.e. $\sim 22.8 nm$) for the Pf$_\gamma$ line and over 24 pixels (i.e. $\sim 34.3 nm$) for Br$_\alpha$. \\
The Hydrogen emission line maps obtained from the corrected and non-corrected data cubes do not differ from each other (the extinction corrected maps are shown in Figs.~\ref{line1mapcorr} and~\ref{line2mapcorr}, respectively). 
This result is expected since the emission lines are not present in the spectrum of the foreground extinction.
Peaks of the Hydrogen emission line maps are either due to the presence of a broad emission line star or to a high emission of the lines. Some peaks may be due to a very low continuum (i.e. with a very low signal to noise ratio) producing artefacts in the emission line maps. These regions are covered by black ellipses in Figs.\ref{line1mapcorr},~\ref{line2mapcorr} and~\ref{line1contline2}. We checked all other peaks and found that except the Wolf-Rayet stars WR1, WR2 and WR3 that we identified in Moultaka et al.~(2005), only one peak is due to the Wolf-Rayet star already identified by Buchholz et al. (2009) located at the south-western corner of the maps in figures \ref{line1mapcorr},~\ref{line2mapcorr} and~\ref{line1contline2}.\\
From these maps, we see clearly that the hydrogen emission lines trace the shape of the mini-spiral. They are obviously produced by the gas in the mini-spiral that is ionized by the hot massive stars. On the other hand, the Pf$_\gamma$ and Br$_\alpha$ line maps appear to be correlated. Also, in Fig.~\ref{line1contline2}, we overlay the contours of the Br$_\alpha$ line map over the Pf$_\gamma$ map where the correlation is clear. This implies that both emissions are produced by the same gas with the same physical conditions. \\

Finally, we find no correlation between the Hydrogen emission line maps and the absorption maps. This implies that the two media of the gas and dust, are not tightly linked. Dust is more confined in the locations of the bright sources while the gas is more spread all over the region. 
Notice that gas is almost absent in the IRS3-IRS7 region while molecular absorption is prominent there.  \\

   \begin{figure}
   
   \includegraphics[width=24pc]{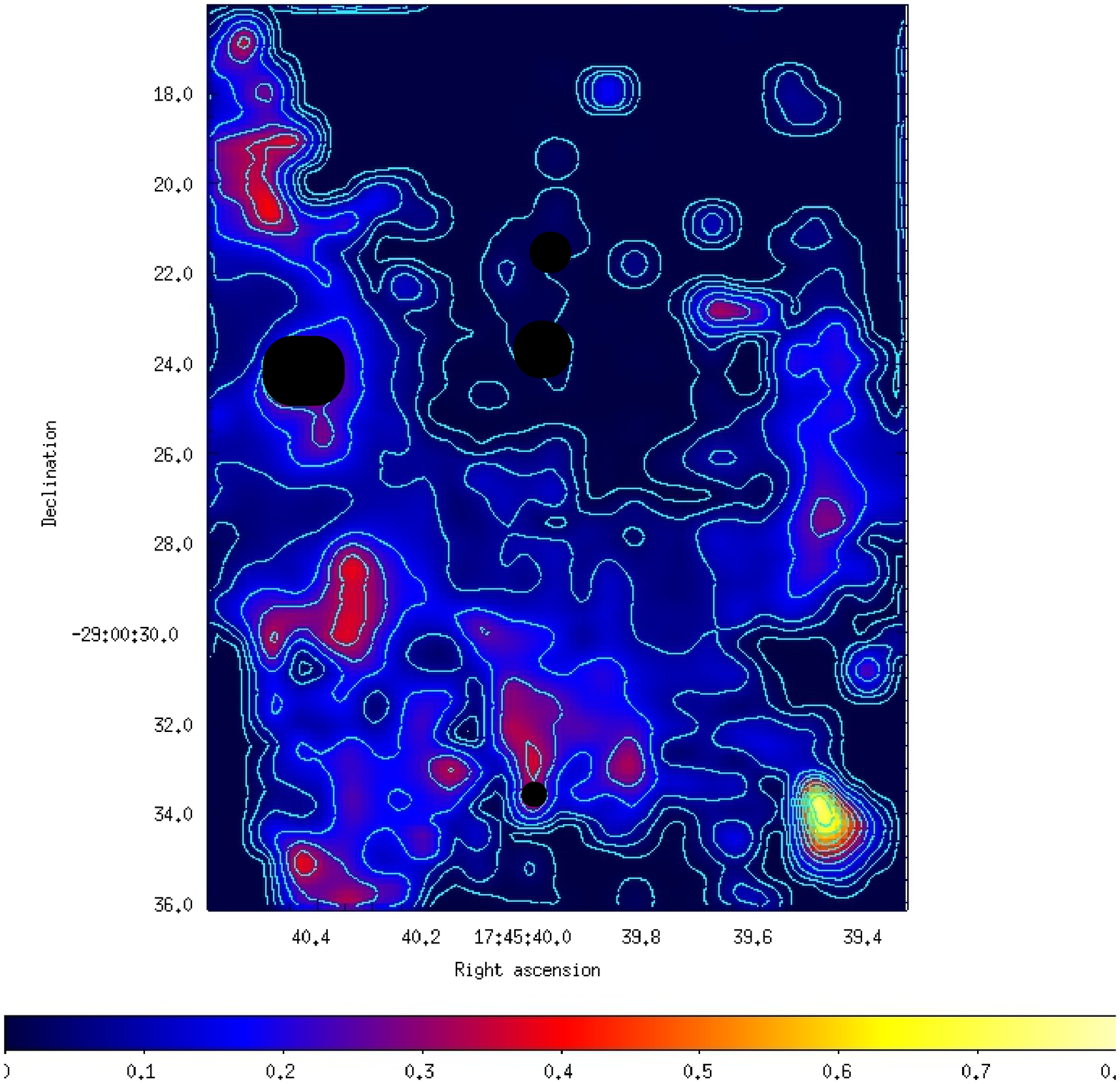}

   \caption{Map of the Pf$_\gamma$ emission line strength corrected for foreground extinction. Contours are also shown for clarity. Contour levels are line strengths = 0.01, 0.04, 0.08  to 0.83 in steps of 0.08. The regions with very low continuum resulting in artificial peaks are covered by black ellipses.}

              \label{line1mapcorr}
    \end{figure}

   \begin{figure}
  
   \includegraphics[width=24pc]{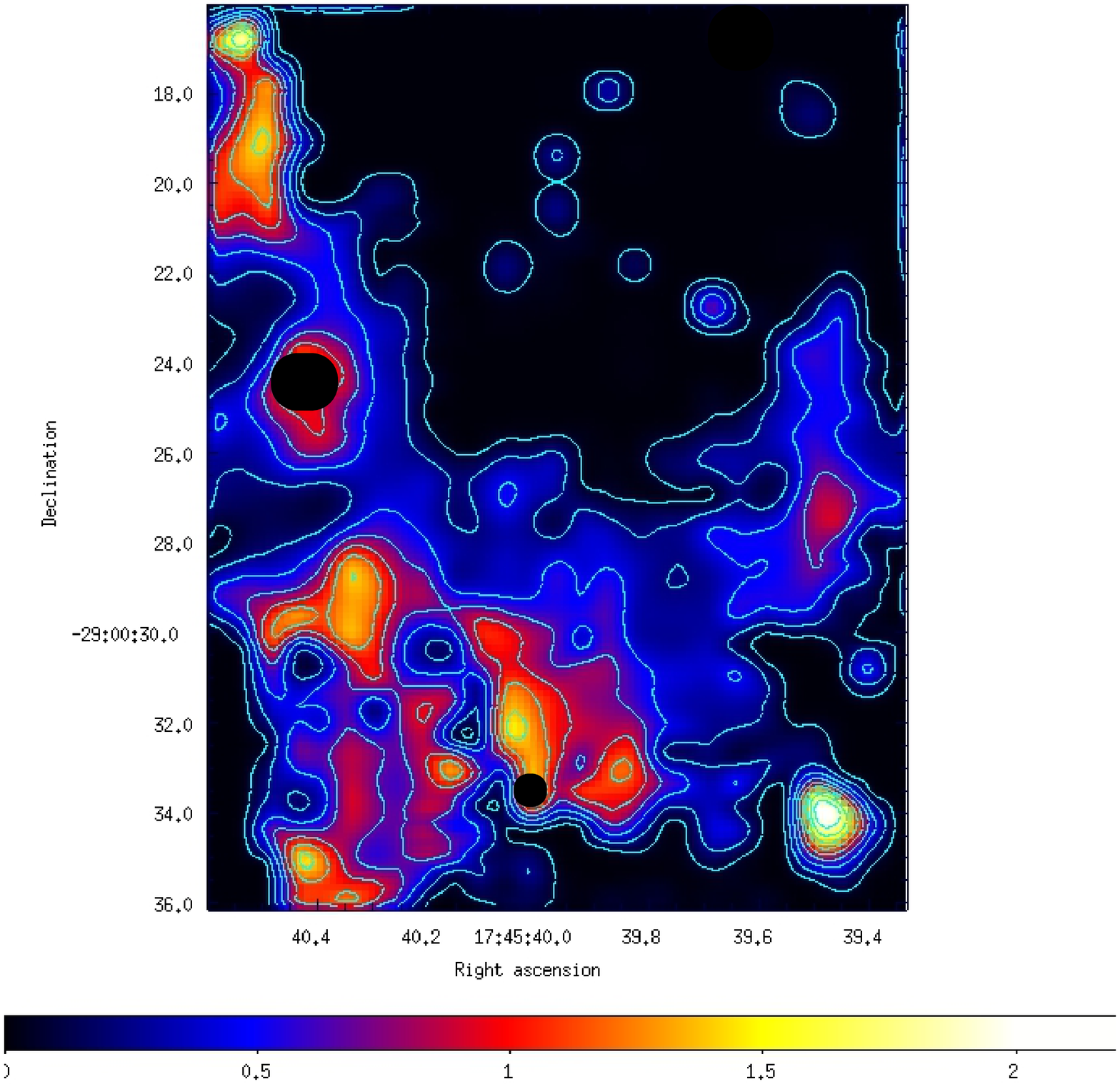}

   \caption{Map of the Br$_\alpha$ emission line strength corrected for foreground extinction. Contours are also shown for clarity. Contour levels are line strengths = 0.07, 0.23 to 2.06 in steps of 0.23. The regions with very low continuum resulting in artificial peaks are covered by black ellipses.}

              \label{line2mapcorr}
    \end{figure}

   \begin{figure}
   
   \includegraphics[width=24pc]{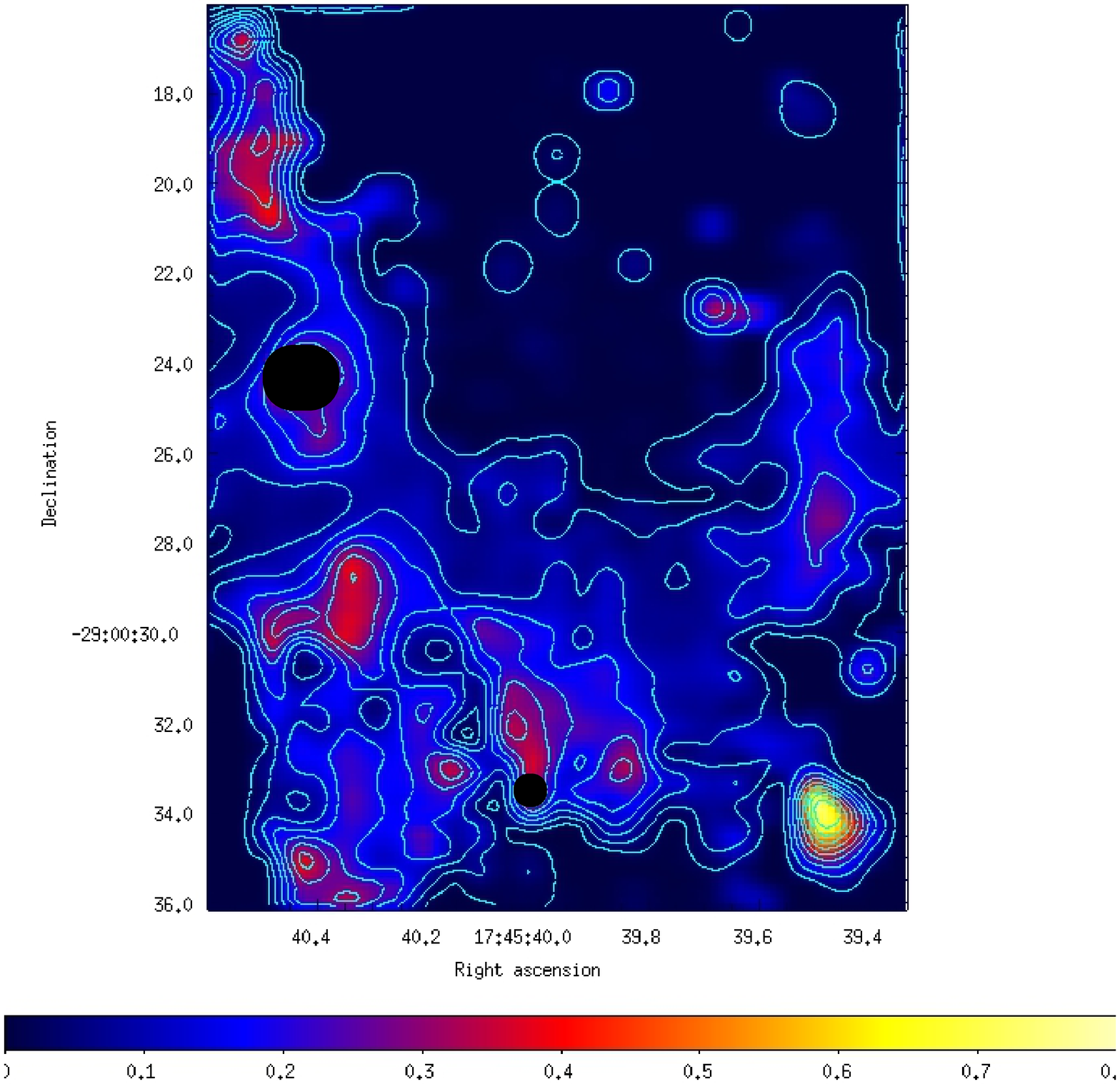}

   \caption{Extinction corrected map of the Pf$_\gamma$ emission line strength with contours of the Br$_\alpha$ emission line strength corrected map. Contour levels are line strengths = 0.07, 0.23 to 2.06 in steps of 0.23. The regions with very low continuum resulting in artificial peaks are covered by black ellipses.}
              \label{line1contline2}
    \end{figure}

\section{Summary and conclusion}

We have produced the first data cube of the central half parsec of our Galaxy in the L-band, using ISAAC spectrograph on the VLT/ESO 8m telescope. This wavelength range is of great interest because it hosts various signatures of the interstellar and the circumstellar medium. In particular, the water ice absorption feature at about 3.0$\mu$m is observed in the dense molecular clouds and especially towards young stellar objects, while the hydrocarbon absorptions at 3.4 and 3.48$\mu$m are commonly observed in the diffuse medium although the 3.48 feature was detected in the spectra of young stellar objects as well (Brooke et al. 1999). Moreover, we observe a number of Hydrogen and Helium recombination lines produced in the external envelopes of young massive stars or in the ionized interstellar medium of the mini-spiral.\\
Using a novel approach to estimate the line of sight absorption due to these features, we corrected the data cube for the foreground extinction and built optical depth maps of the local water ice and hydrocarbon absorptions as well as maps of the local Hydrogen emission line strengths.

In the emission line maps, we found no obvious broad line Wolf-Rayet stars in addition to those previously detected (ex. Paumard et al. 2001, 2006, Tanner et al. 2002, 2005 and Moultaka et al. 2005). This indicates that the census of these stars in the central 0.5 pc must be rather complete.

On the other hand, we found residual absorptions and provided different arguments proving that the residues are clearly produced in the local medium of the central parsec. This finding implies the presence of material at low temperatures - of the order of tens of Kevin - that apparently can survive in the harsh environment of the central stellar cluster close to the central supermassive black hole. This result is also corroborated by our findings of CO ices and cold gaseous CO in the central parsec as described in Moultaka et al. (2015). \\   
We show that the location of the thin dust filaments reported by Mu$\check{z}$i\'c et al. (2007, 2010) are in agreement with the peaks of the optical depths derived from the water ice and hydrocarbon lines. This suggests that the absorbing species are abundant in these high density regions and provide shielding from the harsh interstellar radiation field in the central stellar cluster. In that case the photo-evaporation timescales of 10$^3$ to 10$^5$ years (e.g., Mellema et al. 1998, van Loon \& Oliveira 2003) present a safe lower limit for the evaporation of the water ice and hydrocarbon absorbing clouds.
This helps the compact dusty structures to persist while traveling through the central parsec. It also allows cooler dust/gas aggregates to pass through the central parsec (see Moultaka et al. 2009 and 2015, for more details).
This phenomenon may also be the reason for the good match between the water ice and hydrocarbon absorptions with the dust emission in the 8.6 micron image obtained with VISIR (Viehmann et al. 2006).

{\it This work was supported in part by the Deutsche
Forschungsgemeinschaft (DFG) via SFB~956.
Part of this
work was supported by fruitful discussions with members of
the Czech Science Foundation
DFG collaboration (No. 13-00070J) and with members
of the European Union Seventh Framework Program
(FP7/2007-2013) under grant agreement no 312789; Strong
gravity: Probing Strong Gravity by Black Holes Across the
Range of Masses. }




\clearpage

\appendix
\section{Supplementary figures}
Here we show additional images that help to highlight and comprehend
the analysis presented in the main body of the paper.

\begin{figure}%[h]

\includegraphics[width=25pc]{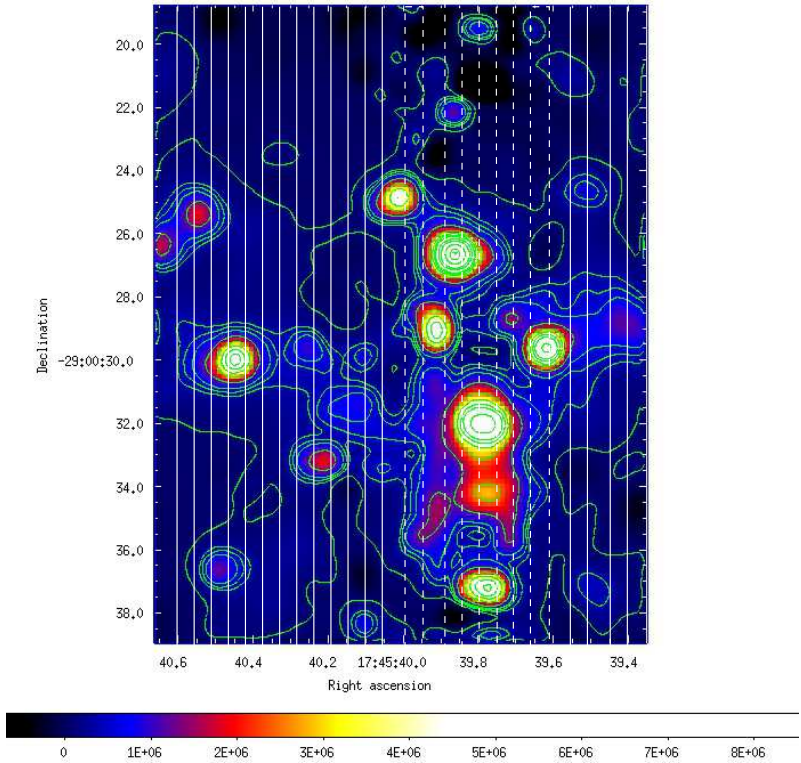} 
\caption{\label{Lband} Integrated emission along the L-band spectral region obtained from our L-band data cube not corrected for the foreground extinction. Contours are also shown for clarity. The surface brightness in L-band is given in contours from 0 to 2.87 Jy/beam in steps of 205 mJy/beam
(Beam= resolution element i.e. FWHM 0.7 arcsec). White lines show the slit positions. The dashed ones correspond to the positions observed in Moultaka et al. (2005).}

\end{figure}

\begin{figure}

\includegraphics[width=24pc]{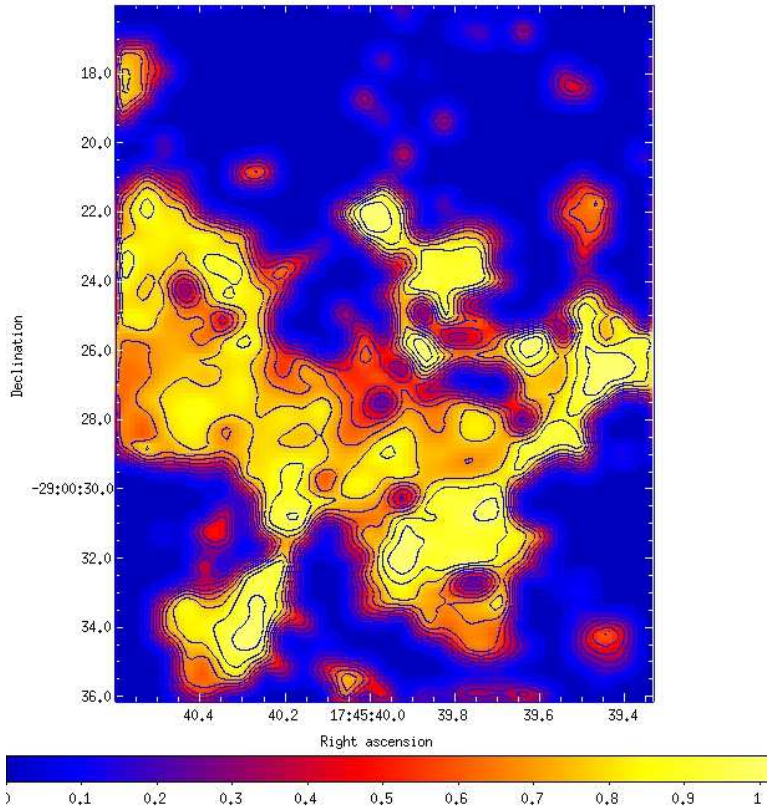} 
\caption{Optical depth map of the water ice absorption feature extracted from our data cube non-corrected for the foreground extinction. Contours are shown for clarity. Contour levels are $\tau$ = 0.03 to 1.06 in steps of 0.09}.

\label{tauIce}
\end{figure}

\begin{figure}

\includegraphics[width=24pc]{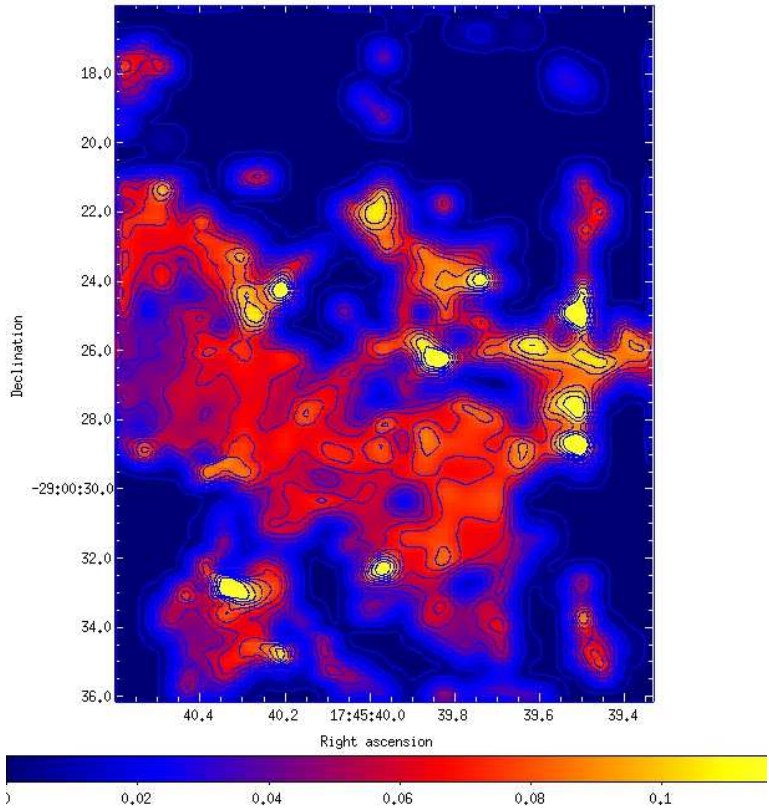} 
\caption{Optical depth map of the hydrocarbon absorption feature extracted from our data cube non-corrected for the foreground extinction. Contours are also shown for clarity. Contour levels are $\tau$ = 0.001, 0.01 to 0.11 in steps of 0.01}.
\label{tauHyd}
\end{figure}
\end{document}